%% file: main-2.tex
\documentclass[letterpaper,dvipsnames]{article}
\pdfoutput=1
\usepackage{jheppub}
\usepackage{amsmath}
\usepackage{graphicx}

\usepackage{array}
\usepackage{ulem}
\usepackage{cancel}
\usepackage{xcolor}
\usepackage{import}

\usepackage{slashed}
\usepackage{afterpage}
\usepackage{appendix}
\usepackage{multirow}
\usepackage{float}
\usepackage{mathtools}
\mathtoolsset{centercolon}

\usepackage{multirow}

\newcommand{\tr}{\text{tr}}

\newcommand{\dd}{\mathrm{d}}

\preprint{CERN-TH-2025-207}

\title{The dark dimension, proton decay, and the length of the M-theory interval}

\author[a]{Mario Reig}

\affiliation[a]{Theoretical Physics Department, CERN, 1211 Geneva 23, Switzerland}

\emailAdd{mario.reig.lopez@cern.ch}

\author[a]{and Ignacio Ruiz}

\emailAdd{ignacio.ruiz.garcia@cern.ch}

\abstract{The existence of a large extra dimension in which only gravity propagates would have spectacular consequences for cosmology and laboratory experiments. In the strong coupling limit of the $E_8\times E_8$ heterotic string theory, the gauge and matter fields live at the end of the eleventh dimension, which becomes a natural candidate for a micron-size \textit{dark dimension}. In this work, however, we show that the length of the M-theory interval is severely constrained by proton decay searches. Our results indicate that in such constructions the size of the eleventh dimension is $R\lesssim \mathcal{O}(10^{-28})$ meters.}

\begin{document}

\maketitle

\section{Introduction}
There is a recent surge in interest in theories with large, purely gravitational dimensions. Motivated by Swampland conjectures~\cite{Ooguri:2006in,Lee:2019wij,Lee:2019xtm,Lust:2019zwm,Obied:2018sgi,Bedroya:2019snp}, a micron-size fifth dimension could be connected to the observed value of the cosmological constant~\cite{Montero:2022prj,Vafa:2024fpx}. This additional dimension would modify Newton's inverse square law at short distances. Current constraints indicate that its size should be $r\lesssim 50$ $\mu$m~\cite{PhysRevLett.124.051301,Lee:2020zjt}. It has also been argued that, in this scenario, a large number of Kaluza-Klein (KK) modes of the graviton~\cite{Gonzalo:2022jac,Obied:2023clp,Law-Smith:2023czn} and other towers in the closed string spectrum~\cite{Langhoff:2025} can be produced via freeze-in after reheating and sizably contribute to the dark matter (DM) of the universe even if the reheating temperature is very low, $T_{\rm RH}\lesssim \mathcal{O}(1)$ GeV. The graviton KK tower can also have important impact in the evolution of stars~\cite{Hardy:2025ajb}. Despite its minimality, concrete UV completions of the dark dimension {remain elusive (see however~\cite{Blumenhagen:2022zzw,Basile:2024lcz,Nian:2024njr})}. 

A general feature of theories with large extra dimensions is that the cut-off of the higher-dimensional theory, above which quantum gravity effects are expected to be relevant, decreases parametrically with respect to the 4-dimensional Planck scale, $M_{\rm Pl,4}$. Historically, this has had applications for the hierarchy problem~\cite{Arkani-Hamed:1998jmv,Arkani-Hamed:1998sfv,Antoniadis:1998ig} (see also~\cite{Dvali:2007wp} for a different realization). In the context of the dark dimension, the existence a single micron-size fifth dimension indicates that the quantum gravity (QG) scale lies around $\Lambda_{\rm QG}\sim 10^{9-10}$ GeV \footnote{The QG scale needs not be the string mass, but rather the fundamental scale of the new theory after decompactifying or going to zero coupling. By the Emergent String Conjecture \cite{Lee:2019xtm,Lee:2019wij} $\Lambda_{\rm QG}$ is expected to be either a higher dimensional Planck mass or the scale of a critical fundamental string. In the strong coupling limit of $E_8\times E_8$ heterotic string theory compactified on a Calabi-Yau threefold $X$ the 4d dilaton remains fixed, with the heterotic string not becoming light in 4d Planck units. On the other hand, the growth of the Ho{\v{r}}ava-Witten interval ~\cite{Horava:1995qa,Horava:1996ma,Witten:1996mz} lowers the 5d and 11d Planck masses, related by $M_{\rm Pl,5}=(V_XM_{\rm Pl,11}^ 6)^{1/3}M_{\rm Pl,11}$, where $V_X$ is the volume of the Calabi-Yau. If $X$ remains fixed in the higher dimensional units, then $M_{\rm Pl,5}$ and $M_{\rm Pl,11}$ scale in the same way, with $M_{\rm Pl,5}\gtrsim M_{\rm Pl,11}$ in the regime $V_X\gtrsim M_{\rm Pl,11}^{-6}$ where higher derivative corrections are suppressed. This numerical difference sets the 11d Planck mass as the $\Lambda_{\rm QG}$.}. Such relatively low QG scale has implications, for example, for our understanding of grand unification. {In~\cite{Heckman:2024trz} the authors argued that unified theories consistent with the dark dimension scenario require the GUT gauge bosons to be solitonic strings of Planckian tension that extend through the dark dimension.} In the context of the QCD axion, the dark dimension can lead to a quite predictive scenario if this particle is localized on the Standard Model (SM) brane, in which case the decay constant lies around $f_a\sim 10^{9-10}$ GeV~\cite{Gendler:2024gdo} (see also~\cite{Li:2024jko}).

A prototypical realization of the dark dimension scenario is a theory with branes that host the SM gauge interactions and fields. These can in principle be localized at a given point in the gravitational dimension or live at the end of a \textit{dark interval} of size $R$. The latter option, as recently discussed in~\cite{Schwarz:2024tet} (see also~\cite{Anchordoqui:2024ajk}), resembles the strong coupling limit of the $E_8\times E_8$ heterotic string theory and suggests that heterotic M-theory with \textit{end-of-the-world} branes hosting an $E_8$ gauge symmetry~\cite{Horava:1995qa,Horava:1996ma,Witten:1996mz} is a good candidate for a UV completion of the dark dimension. While at long distances the theory is effectively five-dimensional~\cite{Lukas:1998yy,Lukas:1998tt}, as the QG scale is approached, the $E_8$ branes are resolved and manifest themselves as 10-dimensional spaces. This 10d space is typically compactified on a Calabi-Yau (CY) threefold as it provides desirable features for low-energy phenomenology such as $\mathcal{N}=1$ supersymmetry\footnote{Our arguments will not depend on this choice, which we take for the sake of simplicity.}. The volume of the CY fixes the localized $E_8$ gauge coupling.

It is well-known that depending on the topological properties of the branes, the size of the M-theory interval can be bounded by consistency arguments~\cite{Witten:1996mz}. A crucial feature to obtain such bound is that, in general, the interval is a warped compact dimension. Because of this warping, when the SM sector lives in the large boundary, one can derive an upper bound to the size of the interval by demanding that the volume of the hidden $E_8$ brane (in the small boundary) does not become negative (see \cite{Cvetic:2024wsj} for a more recent study taking into account non-perturbative corrections). In that case one obtains
\begin{equation}
    R^{\rm max}\sim (\pi|Q|)^{-1}\,,
\end{equation}
where $|Q|\sim \mathcal{O}(1-10^3)\ell_{11}^{-1}$ is the so-called instanton number (to be defined later), and $M_{\rm Pl,11}=(4\pi)^{-1/9}\ell_{11}^{-1}$ is the cut-off of 11d supergravity (SUGRA).

In this work, we show that in cases where the visible sector resides in the small boundary, or even in the absence of warping, the current limit to the proton lifetime~\cite{Super-Kamiokande:2020wjk} bounds the size of the eleventh dimension (which we call $R$) from above. We will show this by quantifying how $M_{\rm Pl,11}$ decreases below the 4d Planck scale, $M_{\rm Pl,4}$, as we increase $R$ for both flat and warped intervals, in the full 11d theory as well as in the effective 5d description. Because consistency of 11d SUGRA requires the KK masses of GUT gauge bosons to be bounded as $M_{\rm KK}\lesssim M_{\rm Pl,11}$, a theory with $M_{\rm Pl,11}\leq 2\times 10^{16}$ GeV leads to excessively fast proton decay\footnote{We also note that SM gauge coupling unification at around $10^{10}$ GeV is incompatible with low-energy measurements of gauge couplings unless the particle content of the SM is drastically modified. See~\cite{Arkani-Hamed:2001gau} for an exotic example.}, {see~\cite{Heckman:2024trz} for a similar idea in the context of general GUT theories}. Our results show that in the most conservative case,
\begin{equation}
    R \lesssim \mathcal{O}(10^{-28})\,\, \text{m} \,,
\end{equation}
providing a robust, model-independent upper bound to the size of the M-theory interval and indicating that it cannot serve as a dark dimension. {We additionally consider an analogous analysis in a more hypothetical setting in which the Standard Model GUT is realized through a stack of M5-branes located at an arbitrary point of the interval and wrapping a curve of the CY, with similar conclusions}. These results are in agreement with the arguments presented in \cite{Heckman:2024trz}, suggesting that if a GUT is realized in nature, the dark dimension requires that it looks very different from standard unified theories where gauge bosons are \textit{point-like} particles up to the QG scale.

\section{Proton decay in heterotic M-theory}
Proton decay requires violation of $U(1)_{B+L}$, a process that is naturally predicted in unified theories~\cite{Georgi:1974sy}. In supersymmetric GUT models, protons are allowed to decay due to dimension-five operators involving superfields, $\int d^2\theta QQQL$, as well as dimension-six operators, $\int d^4\theta Q^2\bar{Q}^\dagger \bar{L}^\dagger$, that typically appear after integrating out heavy GUT gauge bosons transforming as $X_\mu\sim (\mathbf{3},\mathbf{2},-5/6)$~\footnote{{Note that, as usual, the dangerous dimension-4 operators can be avoided by imposing R-partity.}}. The amplitude for the process scales with the gauge boson mass as
\begin{equation}\label{eq:amplitude_p_decay}
    \mathcal{A}\sim \frac{g_{\rm GUT}^2}{M_{X}^2} \,.
\end{equation}
The current results from Super-Kamiokande~\cite{Super-Kamiokande:2020wjk} constrains the proton lifetime to be $\tau_p\gtrsim2.4\times 10^{34}$ years, which corresponds to a GUT scale $M_{\rm GUT}\gtrsim 2\times 10^{16}$ GeV. The bound to the lifetime is expected to be improved by approximately one order of magnitude at Hyper-Kamiokande~\cite{Hyper-Kamiokande:2018ofw}.

{Different symmetry-based mechanisms have been proposed to forbid dimension-5 operators that lead to proton decay in susy GUTs, see for example~\cite{Buchbinder:2013dna,Buchbinder:2014qda} in the context of the $E_8\times E_8$ heterotic string, \cite{Ellis:1997ec} in M-theory, and~\cite{Grimm:2010ez} in F-theory GUTs. Note however that the gauge boson-induced dimension-6 operator arises from the fact that the SM gauge group is embedded into a simple GUT gauge group and cannot be forbidden by symmetry mechanisms. Therefore in susy GUTs the proton decay rate will, at least, scale as $\Gamma_p\sim \alpha_{\rm GUT}^2\frac{m_p^5}{M_{\rm GUT}^4}$. }

In GUT-like string models the proton can decay even if there is no unified theory in the 4d EFT. In heterotic string models proton decay proceeds similar to standard 4d GUT theories~\cite{Witten:1985xc}. Some heterotic models allow to get rid of several of the GUT-like predictions such as some unwanted relations between fermion masses while keeping the unification of gauge couplings~\cite{Green:1987mn}. Similar to gauge coupling unification, in heterotic models compactified on a CY, the predictions for proton decay {from dimension-6 operators} will be nearly unchanged. This occcurs because there always exists a colour triplet gauge boson $X_\mu$ that mediates this process. The exchange of heavy GUT gauge bosons, be it a zero mode that aquires a mass from a GUT Higgs mechanism or a KK mode whose mass is a multiple times the KK scale, $M_X\sim M_{\rm KK}$, is expected to mediate proton decay in a CY where KK number is not conserved. For this reason, up to numerical $\mathcal{O}(1)$ differences that depend on the model details, e.g. the wave functions in the extra dimensions, one expects that the amplitude will scale very similar to the 4d GUT case in~\eqref{eq:amplitude_p_decay}. 

Proton decay has also been studied in GUT-like M-theory models compactified on manifolds of $G_2$ holonomy \cite{Friedmann:2002ty} as well as intersecting D-brane GUT models \cite{Klebanov:2003my,Cvetic:2006iz}, yielding results that are comparable with 4d GUTs up to $\mathcal{O}(1)$ modifications.  In non-unified D-brane models, localization of different quarks and leptons in different places of the gravitational bulk allows to have exponentially suppressed amplitudes~\cite{Arkani-Hamed:1999ylh,Aldazabal:2000cn}, however no such mechanism is available for heterotic models compactified on CY, where matter and gauge fields propagate in 10d. We will assume that in the case of the strongly coupled $E_8\times E_8$ heterotic string the amplitude of the process $p\rightarrow \pi^0e^+$ behaves in a similar way to the perturbative theory, although we contemplate the possibility of $\mathcal{O}(1)$ changes.

\section{Estimates in the full 11d theory\label{sec.11d}}
%


We start with a flux compactification of heterotic M-theory on a warped product $ X\times S^1/\mathbb{Z}_2$, with $X$ being a Calabi-Yau (CY) threefold, that preserves 4d $\mathcal{N}=1$ supersymmetry.  Considering vanishing 4-form flux $G_4$ along the interval direction, $G_{MNPQ11}=0$, the 11d metric takes the form~\cite{Witten:1996mz,Curio:2000dw,Curio:2003ur,Svrcek:2006yi}
\begin{equation}\label{eq. 11d metric}
    \dd s^2_{11}=\hat{G}_{MN}\,\dd x^M\dd x^N=e^{-f(x^{11})}\eta_{\mu\nu}\dd x^\mu \dd x^\nu+e^{f(x^{11})}\left[V_0^{\frac{1}{3}}h_{mn}\dd y^m\dd y^n+R_0^2(\dd x^{11})^2\right]\;,
\end{equation}
with $h_{mn}$ the CY metric and $x^{11}\in[0,\pi]$. Without loss of generality, we will take $\int_X\dd^6y\,\sqrt{h}=1$, so that $V_0$ and $R_0$ denote the volume of $X$ and length of $S^1/\mathbb{Z}_2$, modulo warping. Regarding the expression of $f(x^{11})$, one finds that considering no M5 branes located along the $S^1/\mathbb{Z}_2$ interval
\begin{equation}\label{eq. warping alpha}
    e^{f(x^{11})}=\left(1+Q_0R_0\,x^{11}\right)^{\frac{2}{3}},\quad\text{with }\; Q_i=\frac{1}{32\pi^2\ell_{11}}\int_{X(x^{11}_{i})}\omega\wedge\left[\tr_i(F\wedge F)-\tfrac{1}{2}{\rm tr}(\mathcal{R}\wedge \mathcal{R})\right]\;,
\end{equation}
where $\omega$ is the K\"ahler form on $X$, $F$ and $\mathcal{R}$ are the $E_8\times E_8$ and curvature 2-forms, and $\tr_i$ refers to the trace on the $i$-th boundary, $x^{11}_0=0$ and $x^{11}_1=\pi$. Additionally, anomaly cancellation requires $Q_0+Q_1=0$.\footnote{{ One can consider more involved settings with stacks of space-filling M5-branes perpendicular to the $S^1/\mathbb{Z}^2$ and wrapping some holomorphic 2-cycle $C$ of $X$. This modifies the warped metric \eqref{eq. warping alpha} and the anomaly cancellation condition on the charges. This situation will be studied in more detail in Section \ref{sec. M5 bulk}. For our purposes of this section, where the Standard Model GUT is realized on one of the $E_8$ boundaries, only the $\frac{2}{3}$ exponent and the sign of $Q_0$ are important to bound the possible radius of the $S^1/\mathbb{Z}_2$ interval, so this more involved computations would only amount to a $\mathcal{O}(1)$ difference to our results.
}}
Crucially, depending on the value of the instanton number $Q\equiv Q_0$, the CY volume $V_X(x^{11})=V_X(1+QR_0\, x^{11})^2$ increases or decreases along the interval, with $Q=0$ being the case where we have a flat interval and the compact space is simply the product $X\times S^1/\mathbb{Z}_2$.\footnote{Along this paper we will focus on the case where the 4-form flux vanishes in the $S^1/\mathbb{Z}_2$ direction. For $G_{M\,N\,P\,11}\neq 0$ one finds \cite{Curio:2000dw}
\begin{equation}
    \dd s^2_{11}=\hat{G}_{MN}\,\dd x^M\dd x^N=e^{-\frac{1}{2}f(y^m)}\eta_{\mu\nu}\dd x^\mu \dd x^\nu+e^{f(y^m)}\left[V_0^{\frac{1}{3}}h_{mn}\dd y^m\dd y^n+R_0^2(\dd x^{11})^2\right]\;,
\end{equation}
with the warp factor $e^{f(y^m)}$ depending only on the CY coordinates and not on $x^{11}$, with $\partial_n e^{\frac{1}{2}f(y^m)}=-\frac{\sqrt{2}}{3}G_{nm}{}^m{}_{11}$. This means that the volume of $X$, $V_X=V_0\int_X \dd^6 y\sqrt{h}e^{3f(y^m)}$ does not change along the interval. Dimensional reduction of the EH action results in
\begin{equation}
    \frac{M_{\rm Pl,11}}{M_{\rm Pl,4}}=\left(\frac{V_XR_0}{4\ell _{11}^7}\right)^{-1/2}\;,
\end{equation}
precisely corresponding with the flat regime of our analysis. While embedding the SM with $G_{mnl11}\neq 0$ might be more complicated (the M-theory 3-form propagates along $x^{11}$, which is problematic if we want to localize the SM), the bounds on the size of the interval would be the same as in the flat case with $G_{mnl11}= 0$.
} 

Dimensionally reducing the Einstein-Hilbert term of the 11d action~\cite{Witten:1996mz} we obtain the following relation between the 11d and 4d Planck masses:
\begin{equation}\label{eq. Mpl11}
    \frac{M_{\rm Pl,11}}{M_{\rm Pl,4}}=
    \left\{\frac{3}{32}\frac{V_0R_0}{\ell_{11}^7}\frac{(q+1)^{\frac{8}{3}}-1}{q}\right\}^{-\frac{1}{2}}.
\end{equation}
Here $q$ the dimensionless instanton number defined as
\begin{equation}\label{eq:dimless_q}
    q=\pi R_0Q=\frac{1}{32\pi}\frac{R_0}{\ell{_{11}}}\int_X\omega\wedge\left[\tr(F\wedge F)-\frac{1}{2}{\rm tr}(\mathcal{R}\wedge \mathcal{R})\right]\;.
\end{equation}

In a similar way, dimensionally reducing the Yang-Mills part of the action 
\begin{align}
    S\supset-\frac{1}{8\pi(4\pi\kappa_{11})^{2/3}}\int_{M_{10}^{(i)}}\tr(F_i\wedge\star F_i)\supset-\int\dd^4x\sqrt{-g}\,\frac{1}{2g_{\rm GUT}^2}\tr_i F^2\;,
\end{align}
we obtain the 4d gauge coupling for each of the $E_8$ factors. For simplicity, we assume that the SM is embedded into the same $E_8$ factor. In this case, the 4d gauge coupling of the visible sector is given by
\begin{equation}\label{eq.GUT}
    \alpha_{\rm GUT}=\frac{g_{\rm GUT}^2}{4\pi}=\frac{(4\pi\kappa_{11}^2)^{2/3}}{V_X(x^{11}_i)}=\frac{\ell_{11}^6}{V_0}(1+QR_0x^{11}_i)^{-2}\,,
\end{equation}
with $x^{11}_i=0$ or $\pi$ depending on the boundary hosts the SM. Without loss of generality, we will assume that the visible sector is located in $x^{11}_1=0$, as choosing the opposite $E_8$ amounts to $x^{11}\to\pi-x^{11}$ and $Q\to-Q$ by anomaly cancellation in the absence of additional M5 branes, see comments below \eqref{eq. warping alpha}. 

Using Eqs.~\eqref{eq. Mpl11} and \eqref{eq.GUT}, we obtain a relation between $M_{\rm Pl,11}$, $\alpha_{\rm GUT}$ and the physical length of $S^1/\mathbb{Z}_2$ taking warping into account,
\begin{equation}\label{eq:phys_length}
    R=\pi R_0\frac{3[(1+q)^{4/3}-1]}{4q}\;.
\end{equation}
This relation is given by
\begin{equation}\label{eq. relation}
    \frac{M_{\rm Pl,11}}{M_{\rm Pl,4}}\left(\frac{R}{\ell_{11}}\right)^{\frac{1}{2}}\alpha_{\rm GUT}^{-1/2}=2\sqrt{2\pi
}\left[(1+q)^{4/3}+1\right]^{-1/2}    \,.
\end{equation}
Using the definition of the dimensionless instanton number, 
\begin{equation}
    q=RQ\frac{4q}{3[(1+q)^{4/3}-1]}\,,
\end{equation}
we find that the expression \eqref{eq. relation} is an implicit relation between $M_{\rm Pl,11}$, $\alpha_{\rm GUT}$ and $R$ for $Q\neq 0$. Below we study its implications for the different values of $Q$.

\subsection{The flat case: \texorpdfstring{$Q=0$}{Q=0}}
\label{sec:flat_11d}
We first consider the simplest case, with both $E_8$ branes being topologically identical. In the absence of NS5-branes the Bianchi identity is solved for $\tr_1 F^2 = \tr_2 F^2 = \frac{1}{2}\tr \mathcal{R}^2$. This implies that $Q=q=0$, and from \eqref{eq. relation} we obtain
\begin{equation}\label{eq. relation 0}
    M_{\rm Pl,4}^2=\frac{1}{4\pi}\frac{M_{\rm Pl,11}^2}{\alpha_{\rm GUT}}\left(\frac{R}{\ell_{11}}\right)\;.
\end{equation}
As anticipated above, we find that for $R/\ell_{11}\gg 1$, as expected in the dark dimension scenario~\cite{Montero:2022prj,Vafa:2024fpx}, the 11d SUGRA cutoff $M_{\rm Pl,11}$ becomes parametrically smaller than $M_{\rm Pl,4}$.
Achieving $R\sim \mu \text{m}$ requires a low cut-off scale for 11d SUGRA, $M_{\rm Pl, 11}\approx 10^9$ GeV. As we will see below, however, this is not compatible with the current proton decay bounds. 

In order to avoid excessively fast proton decay we must impose that the GUT gauge bosons mass is $M_{X,Y}\sim g_{\rm GUT}M_{\rm GUT}\lesssim  M_{\rm KK}$, with $M_{\rm GUT}\gtrsim 2\times 10^{16}$ GeV. Together with the bound to the KK scale $M_{\rm KK}\sim V_X(x^{11}_i)^{-1/6}\lesssim M_{\rm Pl,11}$, from~\eqref{eq. relation 0} we find the constraint
\begin{equation}
    \frac{R}{\ell_{11}}\lesssim \left(\frac{M_{\rm Pl,4}}{M_{\rm Pl,11}}\frac{M_{\rm KK}}{M_{\rm GUT}}\right)^2<\left(\frac{M_{\rm Pl,4}}{M_{\rm GUT}}\right)^2\lesssim \mathcal{O}(10^4)\,.
\end{equation}
Imposing that $M_{\rm Pl,11}\gtrsim g_{\rm GUT}M_{\rm GUT}$ and $\alpha_{\rm GUT}=g_{\rm GUT}^2/4\pi\sim 1/25$, this results in an upper bound on the $S^1/\mathbb{Z}_2$ length:
\begin{equation}\label{eq.bound q0}
    \boxed{R\lesssim 1.4\times 10^{-12}{\text{ GeV}^{-1}\approx2.7\times10^{-28}\text{ m}}\,,\,\,\,\text{for }\; Q=0\;}
\end{equation}
This results indicates that the (flat) interval is too small for the dark dimension scenario to be realized.

\subsection{The warped case: \texorpdfstring{$Q\neq 0$}{Q!=0}}\label{sec:warped_11d}
In this section we consider the more phenomenologically interesting case with instanton number $Q\neq 0$ on the $E_8$ brane. In this case $e^{f(x^{11})}$ is non-constant and the volume of the compact CY threefold $X$ varies as we move along the $S^1/\mathbb{Z}_2$.

Historically, the literature has focused in situations where $Q<0$~\cite{Witten:1996mz, Lukas:1998yy,Lukas:1998tt} (see also \cite{Cvetic:2024wsj} for a recent discussion). In this case the SM sector is located at the $x^{11}=0$ brane and the volume of $X$ decreases along the interval. Since $V(x^{11})=V_0(1+QR_0x^{11})^2$, requiring a non-zero CY volume along the whole interval implies that there is a maximal value for the $R_0$ parameter, $R_0^{\rm max}=(\pi|Q|)^{-1}$, being allowed. This results in the upper bound for the physical length of the interval, $R^{\max}=\frac{3}{4}|Q|^{-1}$. Classically, this results in an end-of-the-world brane appearing at finite distance, with spacetime closing at a point and fields not propagating further. Since both $E_8$ branes need to be accessible in order for the compactification to be consistent, this prevents the internal interval from being arbitrarily large. While including quantum corrections regulate the vanishing volume \cite{Witten:1996mz,Cvetic:2024wsj} allowing the $R_0$ field space to be extended beyond the classical boundary, in general the new dual theory is not well understood.\footnote{The authors of \cite{Cvetic:2024wsj} conjecture that this might be M-theory compactified on a curved  7-manifold threaded by some flux, on which a natural realization of a large extra dimension is not obvious.} Because of this, we will only consider $R_0\leq R_0^{\rm max}=(\pi|Q|)^{-1}$. 

Similar arguments to the flat case, where we imposed that $M_{\rm Pl, 11}\gtrsim M_{\rm KK}\gtrsim g_{\rm GUT}M_{\rm GUT}$, result in 
\begin{equation}
   R<\frac{3}{4}\frac{1}{\ell_{11}|Q|}\ell_{11}\lesssim\frac{7\times 10^{-18}\text{ GeV}^{-1}}{\ell_{11}|Q|}\approx\frac{10^{-33}\text{ m}}{\ell_{11}|Q|}\;.
\end{equation}
Since in general we do not expect the (dimensionless) instanton number $\ell_{11}|Q|$ to be arbitrarily small for $Q\neq 0$,\footnote{\label{fn.inst}Expanding the K\"ahler form $\omega=\omega_it^i$ and the second Chern classes $c_2(F)=c_{2,i}(F)\tilde{\omega}^i$, with $\int \omega_i\wedge\tilde{\omega}^j=\delta_i^j$ and K\"ahler moduli $\{t^i\}_{i=1}^{h^{1,1}(X)}$, one can rewrite 
\begin{equation}
    \ell_{11}Q=\frac{1}{4}\int_X\omega\wedge \left[c_2(V)-\frac{1}{2}c_2(TX)\right]=\frac{t^i}{4}\left[c_{2,i}(V)-\frac{1}{2}c_{2,i}(TX)\right]\;.
\end{equation}
Now, since by normalization $\int_X\dd^6y\sqrt{h}=\frac{1}{3!}\kappa_{ijk}t^it^jt^k=1$ (with $\kappa_{ijk}$ the intersection numbers of $X$) and $c_{2,i}(V)\sim \mathcal{O}(1)$ to $\mathcal{O}(10^2)$ and $c_{2,i}(TX)\sim \mathcal{O}(10)$ to $\mathcal{O}(10^{3})$ \cite{Gao:2014nfa,Anderson:2017aux,Berglund:2024zuz}, then we conclude that indeed $\ell_{11}Q\sim \mathcal{O}(1)$ to $\mathcal{O}(10^3)$.
} then indeed one finds 
\begin{equation}\label{eq. bound neg}
    \boxed{R<7\times 10^{-18}\text{ GeV}^{-1}\approx 10^{-33}\text{ m}\,,\quad \text{for  }\;Q<0\;}
\end{equation}
This result indicates that, for $Q<0$, the interval is too small to realize the dark dimension scenario.
\vspace{0.1cm}

\begin{figure}
    \centering
    \includegraphics[width=0.8\linewidth]{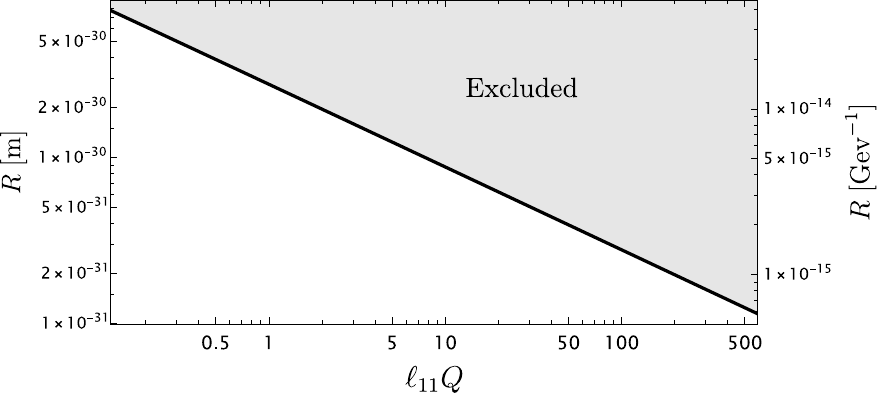}
    \caption{Allowed region for the length of the $S^1/\mathbb{Z}_2$ interval for $Q>0$, as a function of the dimensionless $\ell_{11} Q$ instanton number controlling the warping of the interval. The upper bound comes from imposing that the mass of the KK modes of the GUT gauge bosons is sufficiently heavy to avoid rapid proton decay,  that is imposing $g_{\rm GUT}M_{\rm GUT}\lesssim M_{\rm KK}\lesssim M_{\rm Pl, 11}$ (see text for details). Any region above the solid line is excluded by the current limit from Super-K~\cite{Super-Kamiokande:2020wjk}. For $\ell_{11} Q\to 0^+$ the bound results in the unwarped $Q=0$ case, \eqref{eq.bound q0}.}
    \label{fig.Qpos}
\end{figure}

At first sight, the case $Q>0$ seems more promising to achieve a large eleventh dimension. In this case, the volume of the CY grows along the interval and in principle there is no obstruction to having $R$ arbitrarily large when the SM resides in the small $E_8$ brane, $x^{11}=0$. However, similar to the flat case, existent constraints to the lifetime of the proton will bound the interval from above. To estimate an upper bound to  $R$ we note that
\begin{equation}
   [(1+q)^{4/3}+1]\frac{R}{\ell_{11}}=2\left(\frac{g_{\rm GUT} M_{\rm Pl,4}}{ M_{\rm Pl,11}}\right)^2\leq 2\left(\frac{M_{\rm Pl,4}}{M_{\rm GUT}}\right)^2\approx 3\times 10^4\;,
\end{equation}
with $R/\ell_{11}=\frac{3}{4}[(1+q)^{4/3}-1](\ell_{11}Q)^{-1}$. The above inequality has as solution
\begin{equation}
    \frac{R}{\ell_{11}}\leq\frac{3}{4}\left(\sqrt{1+\frac{8}{3}\left(\frac{M_{\rm Pl,4}}{M_{\rm GUT}}\right)^2\ell_{11}Q}-1\right)(\ell_{11}Q)^{-1}\;,
\end{equation}
which we depict in Figure \ref{fig.Qpos}. The asymptotic behavior reads
\begin{equation}\label{eq:bound_R_with_warping}
R\lesssim\left\{\begin{array}{ll}
    1.4\times 10^{-12}\text{ GeV}^{-1}\approx2.7\times10^{-28}\text{ m} & \text{for }\ell_{11}Q\to 0^+   \\
     \frac{1.4\times 10^{-14}}{\sqrt{\ell_{11}Q}}\; \text{GeV}^{-1}\approx \frac{2.8\times10^{-30}}{\sqrt{\ell_{11}Q}}\; \text{m}& \text{for }\ell_{11}Q\gg1
\end{array}\right.\;,
\end{equation}
recovering the flat $Q=0$ limit and, crucially, showing that $R$ decreases monotonically as the dimensionless instanton number $\ell_{11} Q$ is increased. For generic CY compactifications we expect $\ell_{11} Q\sim \mathcal{O}(1)$ to $\mathcal{O}(10^{3})$ (see Footnote \ref{fn.inst}), but even in special anisotropic cases where $\ell_{11}Q\to 0^+$ one recovers the bound on $R$ from the flat case, \eqref{eq.bound q0}. 

Eq.~\eqref{eq:bound_R_with_warping} illustrates the fact that the bound on $R$ for warped compactifications with $Q>0$ is lower than the flat case. For this reason, we conclude that independent of the presence of warping, avoiding fast proton decay imposes a robust upper bound on the length of the M-theory interval: 
\begin{equation}
    \boxed{R\lesssim 1.4\times 10^{-12}\text{ GeV}^{-1}\approx2.7\times10^{-28}\text{ m}\,,\quad\text{for }\; Q\geq 0\;}
\end{equation}

\section{Estimates in the effective 5d theory}
At intermediate energies $E\ll M_{\rm Pl,11}$ the strongly coupled $E_8\times E_8$ theory looks effectively five-dimensional~\cite{Lukas:1998yy,Lukas:1998tt}, with three-branes living at the end of the fifth dimension. The space-time is in this case $M_4\times S^1/\mathbb{Z}_2$ with a warped interval. These constructions have interesting features for phenomenology, including an attractive solution to the hierarchy problem~\cite{Randall:1999ee} when the visible sector lives in the large boundary. For the sake of completeness, in this section we obtain an upper bound to the size of the interval by using the so-called domain wall solution.

In the regime where the theory is 5-dimensional, the ansatz for the  metric is~\cite{Lukas:1998yy,Lukas:1998tt} (using a notation slightly different than that of \cite{Cvetic:2024wsj}),
\begin{equation}
    \dd s_5^2=e^{2\sigma(y)}\eta_{\mu\nu}\dd x^\mu\dd x^\nu+e^{8\sigma(y)}R_0^2\dd y^2\;,\quad \text{with }\ y\in[0,\pi]\;.
\end{equation}
The warp factor and internal volume of the compact CY (which changes as we move along the interval) can be shown to be
\begin{equation}
    e^{2\sigma(y)}=1+C Q y\;,\qquad V(y)=\frac{4\pi\ell_{11}^6}{g_{\rm GUT}^2}\left(1+CQ y\right)^3\;,
\end{equation}
with $4\pi g_{\rm GUT}^{-2}=V(0)\ell_{11}^{-6}$ as in \eqref{eq.GUT} and $ Q$ a (dimensionful) instanton number defined as in \eqref{eq. warping alpha}. Note that now we find two dynamical parameters, $R_0$ and $C$ (both with units of length) controlling the size of $X$ along the interval and the physical radius of $S^1/\mathbb{Z}_2$:
\begin{equation}
    R=\pi R_0\frac{(1+\pi C Q)^3-1}{3\pi C Q}\;.
\end{equation}
While in the limit $R_0\to\infty$ only the $S^1/\mathbb{Z}_2$ interval decompactifies to 5d, for $C\to \infty$ both the interval and and $X$ become large, resulting in decompactification to the 11d theory.\footnote{\label{fn.qneg5d}Here we have implicitly assumed $Q>0$, since otherwise there is a $C^{\max}=(\pi|Q|)^{-1}$ for which $X$ shrinks to a point in the opposite boundary of the interval. Note however that now $R=\frac{\pi}{3} R_0$, which still can be made large for $R_0\to\infty$.} Reducing the 5d Einstein-Hilbert action down to 4d we obtain the following relation between the two Planck masses:
\begin{equation}
    \frac{M_{\rm Pl,5}}{M_{\rm Pl,4}}=\left[\pi^2\frac{R_0}{\ell_5}\frac{(1+c)^4-1}{c}\right]^{-1/2}=\left[3\pi\frac{R}{\ell _ 5}\frac{(1+c)^4-1}{(1+c)^3-1}\right]^{-1/2}\;,
\end{equation}
with $c=\pi CQ$ being an \textit{dynamical} dimensionless parameter which can change through variations of $C$. Note that for perturbative regimes where the $X$ volume is large in $\ell_{11}$ units we have $M_{\rm Pl,5}>M_{\rm Pl,11}$. Requiring again  that the 11d Planck scale is larger that the GUT scale, this implies the stronger $M_{\rm Pl,5}>M_{\rm GUT}$, which results in
\begin{align}
    R&<\frac{4}{3}\frac{(1+c)^3-1}{(1+c)^4-1}\frac{M_{\rm Pl,4}^2}{M_{\rm GUT}^3}\approx \tfrac{4}{3}\tfrac{(1+c)^3-1}{(1+c)^4-1}\ 7.4\times 10^{-13}\text{ GeV}^{-1}\approx \tfrac{4}{3}\tfrac{(1+c)^3-1}{(1+c)^4-1}\ 1.5\times 10^{-28}\text{ m}\;.
\end{align}
Since $\frac{4}{3}\frac{(1+c)^3-1}{(1+c)^4-1}\leq 1$ for $c\geq0$ the radius $R$ must be lower than the upper bound in the flat $c=0$ case,
\begin{equation}\label{eq.bound pos 5d}
    \boxed{R< 7.4\times 10^{-13}\text{ GeV}^{-1}\approx  1.5\times 10^{-28}\text{ m}\,,\quad \text{for }\ Q\geq 0\;}
\end{equation}
On the other hand, for $Q<0$, where we can have the $X$ threefold shrinking to a point on the opposite boundary and $R=\frac{\pi}{3}R_0$, which now can become large (see Footnote \ref{fn.qneg5d}), we find  
\begin{equation}
    \boxed{R<\frac{4}{3}\frac{M_{\rm Pl,4}^2}{M_{\rm GUT}^3}\sim  10^{-12}\text{ GeV}^{-1}\approx  2\times 10^{-28}\text{ m}\,,\quad \text{for }\ Q< 0\;,}
\end{equation}
recovering bounds of the same order as in the non-negative case \eqref{eq.bound pos 5d} and through the 11d analysis of Section \ref{sec.11d}.

\section{{M5 branes in the bulk}\label{sec. M5 bulk}}
For the sake of completeness, we consider an additional set-up in which the SM is not realized in the boundary of the Ho{\v{r}}ava-Witten interval, but rather on a stack of $n$ M5-branes located a given point $x_{\rm M5}^{11}\in(0,\pi)$ in its interior. In this setting the warping function $e^{f(x^{11})}$ \eqref{eq. warping alpha} for the 11d metric \ref{eq. 11d metric} takes the more general form \cite{Curio:2000dw,Curio:2003ur}
\begin{equation}\label{eq. gen warp}
    e^{f(x^{11})}= \left(1+R_0\sum_i(x^{11}-x_i^{11})\Theta(x^{11}-x_i^{11})Q_i\right)^{2/3}\;,
\end{equation}
with the instantonic charge of the boundaries defined as in \ref{eq. warping alpha}, and that of the M5-brane stacks as
\begin{equation}
    Q_i=\frac{n_i}{4\ell_{11}}\int_{X(x^{11}_i)}\omega\wedge[C]=\frac{n_i}{4\ell_{11}}\frac{{\rm Vol}(C)}{{\rm Vol}(X)^{1/3}}\,,
\end{equation}
where $[C]$ is the 4-form Poincar\'e dual to the 2-cycle $C\subset X$. Anomaly cancellation now requires $\sum_iQ_i=Q(0)+Q(\pi)+\sum_{\rm M5}Q(x^{11}_{\rm M5})=0$. 

Similar as in previous sections, the relation between 4d and 11d Planck masses, as well the physical size of the interval, are given by
    \begin{equation}
        \frac{M_{\rm Pl,11}}{M_{\rm Pl,4}}=\bigg(\frac{3}{32}\frac{R_0V_0}{\ell_{11}^7}\Sigma^{(8)}\bigg)^{-\frac{1}{2}}\,,\quad 
        R=\frac{3\pi}{4}R_0\Sigma^{(4)}\,,
        \end{equation}
where we define
\begin{equation}
    \Sigma^{(a)}=\sum_i\tfrac{\big(1+\sum_{j\leq i}q_j\Delta_{j,i+1}\big)^{a/3}-\big(1+\sum_{j\leq i}q_j\Delta_{j,i}\big)^{a/3}}{\sum_{j\leq i}q_j}>0\,,\quad q_i=\pi R_0Q_i\,,\quad \Delta_{i,j}=\frac{x^{11}_{j}-x^{11}_i}{\pi}\,.
\end{equation}
As discussed in Section \ref{sec:warped_11d}, in order to have a large interval, we take $q_0>0$. Furthermore, in order not to explicitly break SUSY, the stacks have positive charge, $q_i>0$ (otherwise we would be working with $\overline{\text{M5}}$-branes). 

The worldvolume theory of a stack of $n$ M5-branes is given by a $SU(n)$ 6d $\mathcal{N}=(2,0)$ SCFT \cite{Strominger:1995ac,Witten:1996hc}, with a self-dual $SU(n)$-valued $B_2$-field living on the 5-brane. Since the theory is conformal, there is no notion of gauge coupling. While the self-duality of $B_2$ prevents us from writing down a Lagrangian, we can expand
\begin{equation}
    H_3=\dd B_2=\sum_{i=1}^g(\mathcal{F}^i\wedge\omega_i+\bar{\mathcal{F}}^i\wedge\bar{\omega}_i)\,,\quad\text{with }\;\mathcal{F}^i=F^i+i\star_4F^i\,,
\end{equation}
where $\{\omega_i\}_{i=1}^g$ is the set of $g$ holomorphic harmonic 1-forms on the curve $C$, normalized in such a way that over the base of dual 1-cycles $\{A^i,B^i\}_{i=1}^g$ of $H_1(C;\mathbb{Z})$, 
\begin{equation}
    \int_{A^i}\omega_j=\delta_{ij}\,,\qquad\int_{B^i}\omega_i=\Omega_{ij}\,,
\end{equation}
where $\Omega_{ij}$ is the period matrix of $C$, which controls the shape of the curve, rather than its volume. The reduction to 4d then results in 
\begin{equation}
    S_{\rm 4d}\supset\int\Big\{{\rm Im}(\Omega_{ij})\tr(F_i\wedge\star_4 F_j)+{\rm Re}(\Omega_{ij})\tr(F_i\wedge F_j)\Big\}\,,
\end{equation}
where period matrix can then be seen as a \textit{complexified} gauge coupling, $\Omega_{ij}=\frac{\theta_{ij}}{2\pi}+i\frac{4\pi}{g^2_{ij}}$. One can see that, unlike in the case studied in the previous section, where the $E_8$ GUT coupling was a function of the internal volume of the CY, see \eqref{eq.GUT}, here it is independent of any volume, being only a function of the shape of the curve along which we wrap the stack of M5 branes. This implies that $\alpha_{\rm GUT}=\frac{g_{\rm GUT}^2}{4\pi}\sim \tfrac{1}{25}$ now imposes a restriction on the complex structure of the compact manifold (this is, its shape), rather than its size. 

In spite of this, we can still impose the constraint (in order to avoid fast proton decay)
\begin{equation}
    g_{\rm GUT}M_{\rm GUT}\lesssim M_{{\rm KK},C}\lesssim M_{\rm Pl,5}\,,
\end{equation}
where now we only consider the KK modes propagating over the curve $C$ on which the M5-brane stack is wrapped. Using
\begin{equation}
    \frac{M_{\rm Pl,4}}{M_{\rm Pl,5}}\sim\left(\tfrac{R}{\ell_5}\right)^{\frac{1}{2}}=(4\pi)^{-\frac{1}{9}}\left(\tfrac{R}{\ell_{11}}\right)^{\frac{1}{2}}\left(\tfrac{V_0}{\ell_{11}^6}\right)^{\frac{1}{6}}e^{\frac{1}{2}f(x^{11}_{\rm M5})}\,,\quad \frac{M_{{\rm KK},C}}{M_{\rm Pl,4}}\sim\frac{\ell_{11}}{V_0^{1/6}}e^{-\frac{1}{2}f(x^{11}_{\rm M5})}\frac{M_{\rm Pl,11}}{M_{\rm Pl,4}}\,,
\end{equation}
we obtain the following bounds on the sizes of the internal manifolds:
\begin{equation}
        \frac{V_0}{\ell_{\rm 11}^6}\gtrsim \frac{\Sigma^{(4)}}{\Sigma^{(8)}}\,,\qquad \frac{R}{\ell_4}\lesssim\left(\frac{g_{\rm GUT}M_{\rm GUT}}{M_{\rm Pl,4}}\right)^{-3}\left(\frac{\Sigma^{(4)}}{\Sigma^{(8)}}\right)^{-\frac{1}{2}}e^{-\frac{3}{2}f(x^{11}_{\rm M5})}\,,
\    \end{equation}
where for clarity we neglect $\mathcal{O}(1)$ factors. 

\begin{figure}[hbt!]
    \centering
    \resizebox{0.9\textwidth}{!}{%
    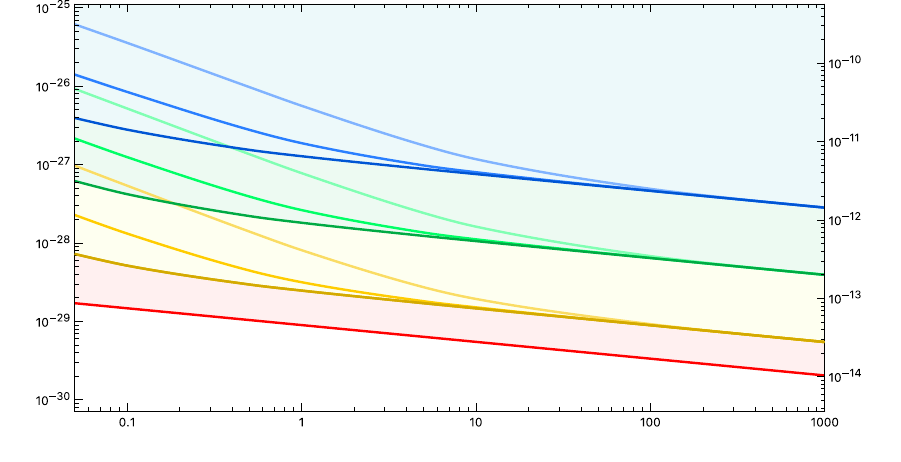
    }
    \caption{Allowed region for the length of the $\mathbb{S}^1/\mathbb{Z}_2$ interval as a function of the dimensionless instanton number $\ell_{11}Q_0$ at the small $E_8$ plane. Different lines correspond to different values of the instanton charge $\ell_{11} Q_{\rm M5}$ and position $x_{\rm M5}^{11}$ of the M5-brane stack. The shaded regions correspond to excluded parameter space after imposing $g_{\rm GUT}M_{\rm GUT}\lesssim M_{{\rm KK},C}\lesssim M_{\rm Pl, 5}$, with $M_{\rm GUT}$ being the lower bound to the GUT scale from proton-decay searches at Super-K~\cite{Super-Kamiokande:2020wjk}, similar to Figure \ref{fig.Qpos}. 
    Note that for large $\ell_{11}Q_0$ the dependence on $\ell_{11} Q_{\rm M5}$ becomes irrelevant. This is also the case when $x_{\rm M5}^{11}=\pi$, i.e, the M5-brane stack is located on top of the large $E_8$-plane.}
    \label{fig.extra}
\end{figure}

The above bound for the size $R$ of the $\mathbb{S}^1/\mathbb{Z}_2$ interval is an implicit expression, since the RHS also depends on $R$ through $q_i=\pi R_0Q_i$, which makes a closed analytic expression not possible. One can numerically obtain the bound on $R$ for different values of the parameters, as we depict in Figure \ref{fig.extra}. Furthermore, in certain limits we recover the following compact expressions
\begin{equation} 
    \frac{R}{\ell_4}\lesssim \left\{\begin{array}{ll}
        \left(\frac{g_{\rm GUT}M_{\rm GUT}}{M_{\rm Pl,4}}\right)^{-9/2}(\ell_{11} Q_0)^{3/4} &\text{if }x_{\rm M5}^{11}=0\,\text{ and }\,Q_0\ell_{11}\to\infty  \\
        \left(\frac{g_{\rm GUT}M_{\rm GUT}}{M_{\rm Pl,4}}\right)^{-18/7}(\ell_{11} Q_0)^{-3/14}(x_{\rm M5}^{11})^{-6/7} & \text{if }x_{\rm M5}^{11}>0,\text{ and }\,Q_0\ell_{11}\to\infty\\
         \left(\frac{g_{\rm GUT}M_{\rm GUT}}{M_{\rm Pl,4}}\right)^{-9/2}(\ell_{11} Q_{\rm M5})^{3/4}&\text{if }\text{if }x^{11}_{\rm M5}<\pi\,\text{ and }Q_0\ell_{11}\to 0\,\\
         \left(\frac{g_{\rm GUT}M_{\rm GUT}}{M_{\rm Pl,4}}\right)^{-3}&\text{if }x^{11}_{\rm M5}=\pi\,\text{ and }Q_0\ell_{11}\to 0\, 
    \end{array}\right.\quad.
\end{equation}

We see that a possible problematic situation occurs for $x_{\rm M5}^{11}=0$, since arbitrarily large $Q_0\ell_{11}$ would allow for an arbitrarily large $R$. However, we note that having the M5-brane stack on top of an $E_8$ plane, be it $x_{\rm M5}^{11}=0$ or $\pi$, can result in a \textit{small instanton transition} \cite{Witten:1996mz,Ganor:1996mu,Seiberg:1996vs}, where the M5-branes, seen as zero-sized instantons of the $E_8$-bundle, are absorbed, in such a way that the total instanton charge of that boundary remains constant. Since this would not allow for a ``stable'' (though SUSY is not broken in this process) realization of the Standard Model through the $SU(n)$ GUT, we will only consider $x_{\rm M5}^{11}>0$.\footnote{As a matter of fact, due to non-perturbative effects caused by Euclidean M2-branes wrapping the ``cylinder'' $[x_i^{11},x_{\rm M5}^{11}]\times C$, with $x_i^{11}=0$ or $\pi$, a non-perturbative potential for the scalar $x_{\rm M5}^{11}$ is generated, see e.g. \cite{Harvey:1999as},
\begin{equation}\label{eq. non-pert pot}
 V_{\rm n.p.}(x^{11}_{\rm M5})\sim \left|\partial_{x_{\rm M5}^{11}}\exp\left(-T_{\rm M2}{\rm vol}(\mathcal{W})+i\int_{\mathcal{W}}C_3\right)\right|^2\sim\left\{\begin{array}{ll}
     \exp\left(-\frac{R_0 V_0^{1/3}}{2\pi^2\ell_{11}^3}x_{\rm M5}^{11}\right) & \text{for }x^{11}_{\rm M5}\sim 0  \\
     -\exp\left[-\frac{R_0 V_0^{1/3}}{2\pi^2\ell_{11}^3}(1+q_0)(\pi-x_{\rm M5}^{11})\right] & \text{for }x^{11}_{\rm M5}\sim \pi
 \end{array}\right.\;,
\end{equation}
so that actually the M5-brane stack would experience a (non-perturbative) repulsive force pulling it away from $x^{11}=0$.
} On the other hand, the instantonic charge of the M5-brane stack, defined as \eqref{eq. gen warp}, has to be finite for our $SU(n)$ GUT purposes, i.e., $n=5$ for $SU(5)$, so an arbitrarily large value for $\ell_{11} Q_{\rm M5}$ is also disfavored, in an argument similar to that of Footnote \ref{fn.inst}. 

We thus conclude that, as shown in Figure \ref{fig.extra}, barring unnatural values for $x^{11}_{\rm\rm M5}$ or $\ell_{11} Q_{\rm M5}$, the size of the Ho{\v{r}}ava-Witten interval is tightly constrained from current bounds on proton decay even when the SM is realized on $M5$-branes. Though the obtained upper bound on the size of $R$ is larger than in the case the GUT gauge group is realized on the boundary $E_8$-planes (compare with Figure \ref{fig.Qpos}), the possible values are still too small to realize the dark dimension scenario. We will close this section noting that, while there is a large abundance of literature constructing the Standard Model on the boundaries of the Ho{\v{r}}ava-Witten interval (see e.g. \cite{Donagi:1999ez,Marchesano:2024gul,Dienes:1996du} for an incomplete list), with stacks of M5-branes in the bulk playing the role of hidden sectors, much less has been studied regarding the possible top-down construction of the Standard Model directly on the M5-brane stacks (see however \cite{Gaiotto:2009hg,Bah:2012dg}, as well as possible connections through the dual F-theory description \cite{Martucci:2015oaa,Assel:2016wcr}, for possible directions).

\section{Conclusions}
Laboratory tests of Newton's inverse square law (see \cite{PhysRevLett.124.051301,Lee:2020zjt} for recent constraints) could reveal the existence of a large extra dimension that modifies gravity at short distances. If such spectacular signal was observed, it would have important implications for some of the currently best understood theories of quantum gravity. Because the existence of a micron-size dark dimension requires a QG scale $\Lambda_{\rm QG}\sim 10^{9-10}$ GeV, string and M-theory completions of the SM that require an effective field theory description in terms of a GUT-like theory, be it in 4d or in 10d, will be generically incompatible with this scenario{, an argument already put forward in~\cite{Heckman:2024trz}}. The reason being that those GUT-like theories with a low QG scale are ruled out by the non-observation of proton decay~\cite{Super-Kamiokande:2020wjk}. 

We have shown this explicitly for the strong coupling limit of the $E_8\times E_8$ heterotic string, obtaining a model-independent upper bound to the size of the eleventh dimension from the current lower bound of the lifetime of the proton.
Our results indicate that the size of the M-theory interval is 
\begin{equation}
    R< 2.7\times 10^{-28}\,\, \text{m}\,.
\end{equation}
This upper bound corresponds to the flat case, see~\eqref{eq.bound q0}. Introducing warping, as shown in Section~\ref{sec:warped_11d}, further decreases this upper bound independently of whether the visible sector lives in the small or the large boundary. {A realization of the Standard Model on a stack fo M5-brane on the bulk of the M-theory interval relaxes the above bound in a couple orders of magnitude but is still too small to accommodate a large extra dimension}. If proton decay is not observed at Hyper-Kamiokande~\cite{Hyper-Kamiokande:2018ofw}, this bound will be strengthened by an $\mathcal{O}(1)$ amount. 

We have focused on heterotic M-theory with 10d end-of-the-world branes compactified on a CY threefold but we expect that {analogous (possibly warped) constructions} will lead to similar conclusions. Our results indicate that in the event of discovering modifications to Newton's gravity at short distances then either non-unified brane models, constructions along the lines of~\cite{Heckman:2024trz} where the GUT gauge bosons are solitonic strings, {F-theory GUT constructions with stacks of 7-branes wrapping holomorphic 4-cycles \cite{Beasley:2008dc,Beasley:2008kw,Marchesano:2022qbx,Marchesano:2024gul}}, or other yet-to-be-discovered non-GUT theories would be the only string completions of the SM compatible with the experiment.

\section*{Acknowledgments}
We thank Salvatore Raucci, John H. Schwarz, Irene Valenzuela and Timo Weigand for useful discussions and illuminating correspondence, as well as work on related projects. The work of I.R. is supported by the European Union through ERC Starting Grant SymQuaG-101163591 StG-2024.

\bibliographystyle{JHEP}
\bibliography{bibliography.bib}

\end{document}

%% file: plot.pdf_tex
\begingroup%
  \makeatletter%
  \providecommand\color[2][]{%
    \errmessage{(Inkscape) Color is used for the text in Inkscape, but the package 'color.sty' is not loaded}%
    \renewcommand\color[2][]{}%
  }%
  \providecommand\transparent[1]{%
    \errmessage{(Inkscape) Transparency is used (non-zero) for the text in Inkscape, but the package 'transparent.sty' is not loaded}%
    \renewcommand\transparent[1]{}%
  }%
  \providecommand\rotatebox[2]{#2}%
  \newcommand*\fsize{\dimexpr\f@size pt\relax}%
  \newcommand*\lineheight[1]{\fontsize{\fsize}{#1\fsize}\selectfont}%
  \ifx\svgwidth\undefined%
    \setlength{\unitlength}{432.35769653bp}%
    \ifx\svgscale\undefined%
      \relax%
    \else%
      \setlength{\unitlength}{\unitlength * \real{\svgscale}}%
    \fi%
  \else%
    \setlength{\unitlength}{\svgwidth}%
  \fi%
  \global\let\svgwidth\undefined%
  \global\let\svgscale\undefined%
  \makeatother%
  \begin{picture}(1,0.51261313)%
    \lineheight{1}%
    \setlength\tabcolsep{0pt}%
    \put(0,0){\includegraphics[width=\unitlength,page=1]{plot.pdf}}%
    \put(0.55761041,0.40724465){\color[rgb]{0,0,0}\makebox(0,0)[lt]{\lineheight{1.25}\smash{\begin{tabular}[t]{l}\LARGE{Excluded}\end{tabular}}}}%
    \put(0.77392087,0.11942483){\color[rgb]{0.50196078,0,0}\rotatebox{-5.6794421}{\makebox(0,0)[lt]{\lineheight{1.25}\smash{\begin{tabular}[t]{l}$x_{\rm M5}^{11}=\pi$\end{tabular}}}}}%
    \put(0.77392098,0.15610077){\color[rgb]{0.4,0.4,0}\rotatebox{-5.6794421}{\makebox(0,0)[lt]{\lineheight{1.25}\smash{\begin{tabular}[t]{l}$x_{\rm M5}^{11}=1$\end{tabular}}}}}%
    \put(0.77392098,0.23091452){\color[rgb]{0,0.50196078,0}\rotatebox{-5.6794421}{\makebox(0,0)[lt]{\lineheight{1.25}\smash{\begin{tabular}[t]{l}$x_{\rm M5}^{11}=0.1$\end{tabular}}}}}%
    \put(0.77392094,0.30532499){\color[rgb]{0,0,0.50196078}\rotatebox{-5.6794421}{\makebox(0,0)[lt]{\lineheight{1.25}\smash{\begin{tabular}[t]{l}$x_{\rm M5}^{11}=0.01$\end{tabular}}}}}%
    \put(0.47450191,0.0090937){\color[rgb]{0,0,0}\makebox(0,0)[lt]{\lineheight{1.25}\smash{\begin{tabular}[t]{l}\Large{$\ell_{11} Q_{0}$}\end{tabular}}}}%
    \put(0.02537541,0.25161194){\color[rgb]{0,0,0}\rotatebox{90}{\makebox(0,0)[lt]{\lineheight{1.25}\smash{\begin{tabular}[t]{l}\Large$ R\,[\text{m}]$\end{tabular}}}}}%
    \put(0.98825663,0.22483717){\color[rgb]{0,0,0}\rotatebox{90}{\makebox(0,0)[lt]{\lineheight{1.25}\smash{\begin{tabular}[t]{l}\Large$R\,[\text{GeV}^{-1}]$\end{tabular}}}}}%
    \put(0.1940542,0.18925377){\color[rgb]{0.4,0.4,0}\rotatebox{-6.824548}{\makebox(0,0)[lt]{\lineheight{1.25}\smash{\begin{tabular}[t]{l}$\ell_{11}Q_{\rm M5}=0.1$\end{tabular}}}}}%
    \put(0.30336729,0.21234615){\color[rgb]{0.4,0.4,0}\rotatebox{-10.649519}{\makebox(0,0)[lt]{\lineheight{1.25}\smash{\begin{tabular}[t]{l}$\ell_{11}Q_{\rm M5}=1$\end{tabular}}}}}%
    \put(0.42408779,0.21453812){\color[rgb]{0.4,0.4,0}\rotatebox{-14.027175}{\makebox(0,0)[lt]{\lineheight{1.25}\smash{\begin{tabular}[t]{l}$\ell_{11}Q_{\rm M5}=10$\end{tabular}}}}}%
    \put(0.15278894,0.27239149){\color[rgb]{0,0.50196078,0}\rotatebox{-7.3283527}{\makebox(0,0)[lt]{\lineheight{1.25}\smash{\begin{tabular}[t]{l}$\ell_{11}Q_{\rm M5}=0.1$\end{tabular}}}}}%
    \put(0.29782932,0.29253957){\color[rgb]{0,0.50196078,0}\rotatebox{-10.502654}{\makebox(0,0)[lt]{\lineheight{1.25}\smash{\begin{tabular}[t]{l}$\ell_{11}Q_{\rm M5}=1$\end{tabular}}}}}%
    \put(0.4190989,0.30174352){\color[rgb]{0,0.50196078,0}\rotatebox{-16.289678}{\makebox(0,0)[lt]{\lineheight{1.25}\smash{\begin{tabular}[t]{l}$\ell_{11}Q_{\rm M5}=10$\end{tabular}}}}}%
    \put(0.16119686,0.3425047){\color[rgb]{0,0,0.50196078}\rotatebox{-7.5079731}{\makebox(0,0)[lt]{\lineheight{1.25}\smash{\begin{tabular}[t]{l}$\ell_{11}Q_{\rm M5}=0.1$\end{tabular}}}}}%
    \put(0.25648865,0.38356148){\color[rgb]{0,0,0.50196078}\rotatebox{-14.191486}{\makebox(0,0)[lt]{\lineheight{1.25}\smash{\begin{tabular}[t]{l}$\ell_{11}Q_{\rm M5}=1$\end{tabular}}}}}%
    \put(0.33287355,0.4093844){\color[rgb]{0,0,0.50196078}\rotatebox{-20.138211}{\makebox(0,0)[lt]{\lineheight{1.25}\smash{\begin{tabular}[t]{l}$\ell_{11}Q_{\rm M5}=10$\end{tabular}}}}}%
  \end{picture}%
\endgroup%